\documentclass[9pt,twocolumn,twoside]{pnas-new}

\templatetype{pnasresearcharticle} 

\newcommand{\tmop}[1]{\ensuremath{\operatorname{#1}}}

\newcommand{\ds}{\displaystyle}
\usepackage{xcolor}
\definecolor{red}{rgb}{1,0,0}
\definecolor{blue}{rgb}{0,0,1}

\graphicspath{{../Figs/}}

\title{Dimple drainage before the coalescence of a droplet deposited on a smooth substrate}

\author[a]{Laurent Duchemin}
\author[b]{Christophe Josserand}
\affil[a]{PMMH, CNRS, ESPCI Paris, Universit{\'e} PSL, Sorbonne
Universit{\'e}, Universit{\'e} de Paris, F-75005, Paris, France 
and Aix Marseille Univ, CNRS, Centrale Marseille, IRPHE, Marseille, France}

\affil[b]{LadHyX, UMR 7646 CNRS \& Ecole Polytechnique, IP Paris, Route de Saclay, 91128 Palaiseau, France}

\leadauthor{Duchemin}

\significancestatement{When a millimetric droplet is laying on a smooth surface, either liquid or solid, the slow gravity drainage of the air layer separating the surface from the drop leads to the entrapment of a thin bubble underneath. In this work, we investigate theoretically, and through numerical computations, the bubble drainage dynamics on a solid and a liquid film. We show and explain how this dynamics before coalescence is much faster on a thin liquid film than on a solid surface.}

\authorcontributions{L.D. and C.J. designed research; L.D. and C.J. performed research; L.D. and C.J. analyzed data; and L.D. and C.J. wrote the paper.}
\authordeclaration{The authors declare no conflict of interest.}
\correspondingauthor{\textsuperscript{2}To whom correspondence should be addressed. E-mail: laurent.duchemin@espci.fr}

\keywords{Drop $|$ Coalescence $|$ Film $|$ Drainage} 

\begin{abstract}
  Thin liquid or gas films are everywhere in nature, from foams to sub-millimetric bubbles at a free surface, and their rupture leaves a collection of small drops and bubbles. However, the mechanisms at play responsible for the bursting of these films is still in debate. The present study thus aims at understanding the drainage dynamics of the thin air film squeezed by gravity between a millimetric droplet and a smooth solid or a liquid thin film. Solving coupled lubrication equations and analyzing the dominant terms in the solid and liquid-film cases, we explain why the drainage is much faster in the liquid-film case, leading often to a shorter coalescence time, as observed in recent experiments.
\end{abstract}

\dates{This manuscript was compiled on \today}
\doi{\url{www.pnas.org/cgi/doi/10.1073/pnas.XXXXXXXXXX}}

\begin{document}

\maketitle
\thispagestyle{firststyle}
\ifthenelse{\boolean{shortarticle}}{\ifthenelse{\boolean{singlecolumn}}{\abscontentformatted}{\abscontent}}{}

\dropcap{C}oalescence is a major process in multiphase flows since it controls, with the break-up dynamics, the global evolution of the number of droplets and bubbles in the flow. It is therefore crucial for spray cooling and coating~\cite{ristenpart2006coalescence}, viscous flows in capillaries~\cite{lamstaes,Li2019}, emulsion and foams~\cite{breward2002drainage,zapryanov1983emulsion}, or droplet transmission of diseases~\cite{Bourouiba2014,Bourouiba2019} for instance. Coalescence is in fact a complex phenomenon, the effective contact between two liquid bodies being not at all automatic as the interfaces approach each other, since surface interactions and viscous forces can in particular contribute to separate the interfaces~\cite{neitzel2002noncoalescence,janssen2006axisymmetric,davis1989lubrication,klaseboer2000film}. These forces can lead to the striking ``floating drops'' phenomena, as already mentioned by Lord Rayleigh~\cite{Rayleigh1899} and Osborne Reynolds~\cite{reynolds1881drops}: for instance, it can be observed in an everyday life ``experiment'' by watching small coffee drops dancing in the mug above the liquid surface before sinking\cite{amarouchene2001noncoalescing}, or when a drop is deposited on a vibrating liquid bath\cite{couder2005bouncing,couder}. The coalescence dynamics is controlled at first by the drainage of the thin fluid film separating the two liquid bodies, whether we consider drop-drop, drop-film or drop-substrate coalescences. The drainage of this separating film determines the time for coalescence that can be defined as the time between the beginning of the drainage regime (where no motion is present but the drainage) to the time where the film ruptures so that the two liquid bodies enter in contact, or the liquid starts to wet the substrate.
Such thin film drainage follows a lubrication dynamics~\cite{chan2011film,yiantsios1990,yiantsios1991close,frankel1962dimpling,hartland1977model,hartland1994dimple,lin1982thinning}, where the viscosity of the surrounding fluid induces high pressure in the film that leads to the interface deformation and sometimes to the entrapment of an air bubble at coalescence as observed in drop impacts for instance~\cite{Thoroddsen2005,Thoroddsen2012b}. 
However, the lubrication dynamics prevents mathematically the rupture of the film in finite time (at least when surface tension is present\cite{DucheminJosserand2011}) even when the film is squeezed, so that the final stage of coalescence has to involve additional physical mechanisms, dominant at microscopic or even nanoscopic scales. Usual suspects are van der Waals interactions~\cite{zhang1999similarity}, surface roughness, thermal fluctuations, Marangoni currents and non-continuum effects~\cite{lhuissier2012bursting}, although their precise implications remain largely an open question.
Even if this final stage of the coalescence dynamics can exhibit thus high fluctuations, it can usually be described as a thickness cut-off below which the film rupture happens rapidly, so that the lubrication dynamics can be taken as the dominant mechanism to determine the coalescence time. Eventually, even the drainage time can exhibit large variations in experiments and models, mostly poorly understood, so that it is crucial to have a better comprehension of the lubrication dynamics of drainage in the different configurations encountered~\cite{klaseboer2000film,lo2017}.

Drop coalescence~\cite{kavehpour2015coalescence,eggers1999coalescence,yoon2007coalescence} can be obtained through mainly two different practical configurations: drop collision (to an other drop, a bath, a substrate in particular) where the drop velocity is the main control parameter~\cite{JieLi2016}; drop deposition, the situation considered in the present paper, where the drop is smoothly deposited on the substrate, meaning that its velocity is zero and that the weight of the drop only is leading to its coalescence with the substrate~\cite{lo2017}. 
While most of the experimental, numerical and theoretical studies have focused on the drop/drop collision, only little attention has been paid eventually to the case of the deposition of a drop on a smooth liquid substrate, where the coalescence dynamics is almost quasi-static since it is driven by the weight of the drop that squeezes the air gap between the drop and the wetted substrate. 
Recently, an experimental investigation of the coalescence of a millimetric drop gently laying on a thin viscous film \cite{lo2017} has revealed unexpected drainage dynamics: the drainage time is seen much shorter in the presence of a very thin viscous film than the time of touch down estimated through simple lubrication arguments.
The authors show experimentally that tangential flows at both interfaces between the gas and the liquids cannot be neglected as it is often done in theory, allowing a more rapid drainage of the gas layer~\cite{jones1978film}. These tangential flows are created by the viscous entrainment of the liquids by the gas layer that is squeezed, a mechanism also observed recently in Leidenfrost configuration~\cite{Ambre2018}. In order to elucidate the influence of these tangential flows in the drainage dynamics, the
present study investigates the drainage dynamics of the thin air film squeezed between a weighting millimetric drop and a solid or wetted surface, varying thus the flow structure of the interstitial gas layer.

\subsection*{Problem formulation}

We consider a liquid droplet landing gently on a wetted surface. 
The main question we want to address is thus: how long does it take for the two liquid surfaces to reach a threshold thickness (defined by the rupture mechanism) as the viscosity of the liquid film varies? 

\begin{figure}[h]
  \includegraphics[width=8.7cm]{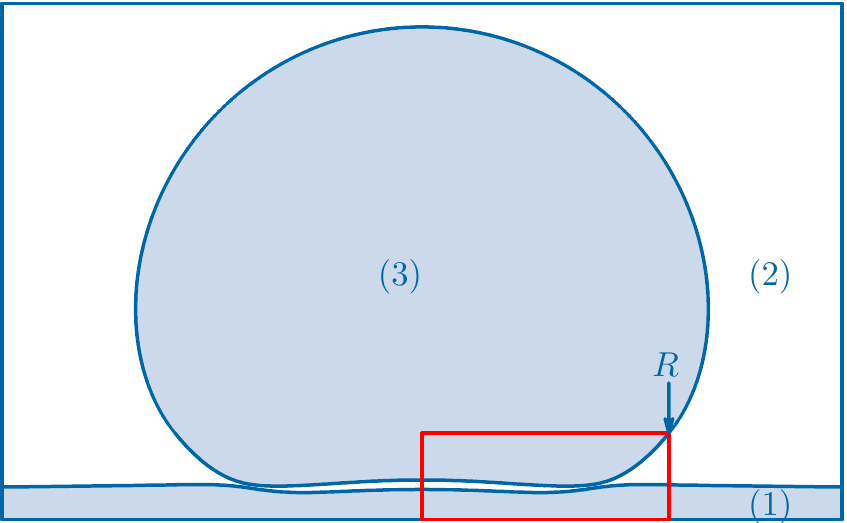}\\
  \\
  \includegraphics[width=8.7cm]{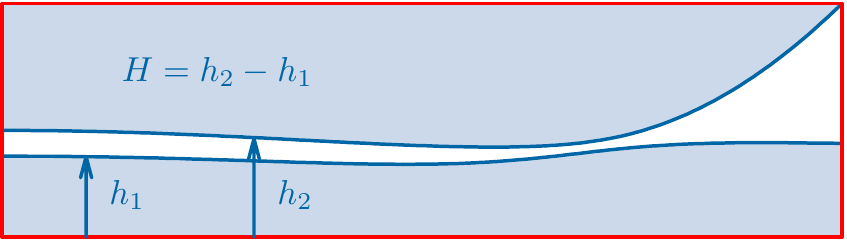}
  \caption{Shape of a droplet lying on a thin liquid film. Unerneath the drop, a gas dimple is entrapped and the liquid film is deformed. The computational domain is defined as the radial extent $[0,R]$, where $R$ is the radius at which the slope of the droplet interface equals $1$. The top surface of the drop corresponds to a static equilibrium shape, under the effects of gravity and surface tension.\label{sketch}}
\end{figure}

The liquid film height is denoted by $h_1 (r,t)$, and the local height of the droplet interface is $h_2 (r, t)$, where $r$ is the radial coordinate according to the axis of symmetry, and $t$ is time. As seen in figure \ref{sketch}, the three domains (liquid film,
gas film, liquid droplet) are denoted respectively $(1)$, $(2)$ and $(3),$ and
every physical parameter (density, viscosity, etc...) will be noted with an
index corresponding to the relevant region.

The drop is weighting on the thin gas and liquid films, creating a pressure field that exactly compensates the drops weight (inertia can be neglected in the drainage regime). 
Viscous and pressure forces balance inside each film creating viscous flows leading to their squeezing. 
We model the dynamics using thus the lubrication equation both in the gas and liquid films (domains 1 and 2 in figure \ref{sketch}) while a free slip boundary condition is taken at the interface between the drop and the gas film~\cite{yiantsios1991close,davisRPM}. 
Indeed, considering a liquid motion inside the drop on the scale of the radius of the drop $a$, as experimentally reported in \cite{lo2017}, the tangential stress continuity across the gas film (of typical thickness $\bar{H}$) and the drop allows to neglect the shear stress at the interface on the gas side as long as $(\eta_2/\eta_3)(a/\bar{H})\gg 1 $, which is the case in the experiments \cite{lo2017}.

The weight of the drop is transmitted to the films through the interface normal stress condition, involving a pressure jump
because of the surface tension.
In the present study, for the sake of simplicity of the equations, we consider the two-dimensional version of the dynamics. It corresponds to assuming that axi-symmetric terms can be neglected, which is acceptable as soon as $\partial A / \partial r \gg A / r$, for any field $A (r, t)$, an approximation that is valid far from the symmetry axis, and thus particularly near the neck
separating the entrapped gas bubble from the surrounding, as argued for instance in \cite{LisDuc08} in a similar context. Finally, in the lubrication approximation, we take the linearized curvatures for
small slopes ($|\partial_r h| \ll 1$)~:
\[ \kappa_1 \simeq \frac{\partial^2 h_1}{\partial r^2} \qquad \tmop{and}
   \qquad \kappa_2 \simeq \frac{\partial^2 h_2}{\partial r^2} \]

Using lubrication theory in the two films, matching tangential velocities
and shear stress between regions 1 and 2, free slip boundary condition between regions 2 and 3, we obtain two coupled evolution equations for $h_1$ and $H
\equiv h_2 - h_1$:

{\begin{equation}
  {\frac{\partial h_1}{\partial t} =  
  \frac{\partial}{\partial r} \left( \frac{h_1^3}{3 \eta_1} \frac{\partial
  p_1}{\partial r} \right. + \left. \frac{h_1^2 H}{2 \eta_1} \frac{\partial
  p_2}{\partial r}  \right) \label{lub1}}
\end{equation}

\begin{equation}
  {\frac{\partial H}{\partial t} = 
  \frac{\partial}{\partial r} \left( \frac{H^3}{3 \eta_2} \frac{\partial
  p_2}{\partial r} \right. + \left. \frac{h_1 H}{2 \eta_1}  \left\{ h_1
  \frac{\partial p_1}{\partial r} + 2 H \frac{\partial p_2}{\partial r}
  \right\}  \right)} \label{lub2}
\end{equation}}

For each equation, one recognizes the usual lubrication term in the first term of the right hand side, while the second term corresponds to the coupling between the two layers due to the tangential velocity boundary conditions. 
While in both films the gravity can be neglected~\cite{yiantsios1990}, the coupling with the 
weighting drop is made through the pressure fields. Assuming small capillary numbers in the drop (in the experiments of \cite{lo2017}, the capillary numbers are always smaller than one, and mostly below $0.1$), the pressure inside the drop can be taken purely hydrostatic, so that
the pressure in the gas film (constant in $z$ in the lubrication approximation) is given by Laplace's law :
\begin{equation}
  p_2 = p_3 (h_2) - \gamma \kappa_2 \simeq p_0 - \rho_3 g h_2 - \gamma \frac{\partial^2 h_2}{\partial r^2},
  \label{lap2}
\end{equation}
where $p_0$ is a reference pressure to be determined by the free surface boundary condition at the top of the drop, $\gamma$ the surface tension. 
The pressure in the liquid film (region 1) is also given by Laplace's law, yielding:
\begin{equation}
  p_2 - p_1 (h_1) = \gamma \kappa_1 \simeq \gamma \frac{\partial^2 h_1}{\partial r^2}.
 \label{lap1}
\end{equation}
The fact that, within these approximations, we can neglect the influence of the velocity field in the drop for the pressure that 
is thus purely hydrostatic, indicates that the deformation of the drop is driven only by the gravity field and the lubrication pressure beneath the drop. Therefore, the upper drop shape is simply that of a sessile drop, balancing surface tension with 
gravity. It means that, beside the region where the gas film dynamics is relevant, the drop shape is static! This remark is
crucial for solving numerically the system of equations above Eqs. (\ref{lub1}), (\ref{lub2}), (\ref{lap2}) and (\ref{lap1}): as shown on figure \ref{sketch}, we introduce a distance $R$, large enough so 
that the lubrication pressure can be neglected. Actually, $R$ is chosen such that the slope of the sessile top surface that patches the film interface at $r=R$ is of the order of unity (in the numerics shown here we have taken it equal to $1$ and we have checked that the results are mostly unchanged when varying this parameter around this value), see figure \ref{sketch}.  The films equations (\ref{lub1},\ref{lub2}) are solved for $r<R$ while the drop shape is that of the static one outside this region (see more details in the Materials and Methods section).

The system of equations is controlled by the following dimensionless numbers, using for the typical radius of the drop, the value of $R$ computed using the slope $1$ for the patching condition between the sessile drop and the gas layer/drop interface:
$$
     \tmop{Bo}  =  \ds\frac{\rho_3 g R^2}{\gamma}, \quad
     \tmop{St}_1 = \ds\frac{\eta_1}{\rho_3 R \sqrt{g R}}, \quad
     \tmop{St}_2 = \ds\frac{\eta_2}{\rho_3 R \sqrt{g R}}.
$$
The last dimensionless number is the ratio between the asymptotic thickness of liquid film $h_0 = \lim_{r \rightarrow \infty} \; h_1 (r)$ and $R$~:
$$ \alpha=h_0/R.$$ 
In the present study, the Bond number $\tmop{Bo}$ is
assumed to have a moderate value, neither too big nor too small compared to
one ($\tmop{Bo} = \mathcal{O} (1)$). Notice that the experiments described in {\cite{lo2017}} correspond to significantly smaller droplets, with $\tmop{Bo} \simeq 0.36$ based on the equivalent radius of the droplet. 
 In the following, we shall investigate the drainage dynamics as the parameters of the liquid film change, through the variation of our control parameter $\tmop{St}_1$. To dimensionalize the other parameters, we shall use typical values of the Stokes numbers
from the experiments described in {\cite{lo2017}} : $\tmop{St}_2 \simeq 1.25 \times 10^{- 4}$ and $\alpha \simeq 0.02$.
With these parameters, and after computing the top sessile drop, the value of $R$ for all the computations is $R\simeq 1.4$~mm. 
We will thus vary the liquid film parameters, from the typical values of the experiments $\tmop{St}_1 \simeq 0.6$ towards the deposition on a solid substrate that can be accounted for simply in our equations by taking the limit $\tmop{St}_1\rightarrow \infty$.

\begin{figure}[h]
  \includegraphics[width=8.7cm]{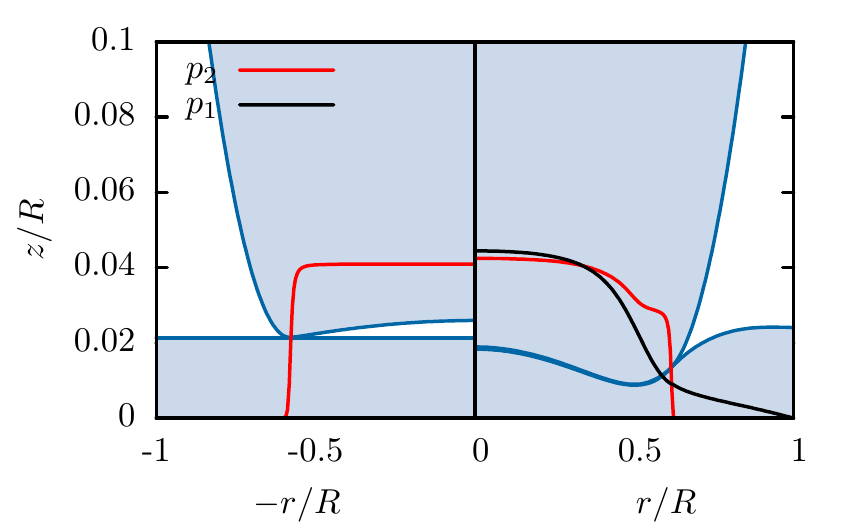}
  \caption{Dimensionless droplet and film shapes $h_2/R$ and $h_1/R$ at the dimensionless time $\tau=t\sqrt{g/R} = 360$, in the case of a solid
  surface (Left, $\tmop{St}_1 = \infty$) and a liquid film (Right,
  $\tmop{St}_1 \simeq 0.6$). Parameters are : $\tmop{Bo} = 1.01$, $\tmop{St}_2
  = 1.25 \times 10^{- 4}$. The red curve is the pressure in the gas film, and
  the black curve the pressure in the liquid film (pressure scale not indicated). At the pinch-point, the
  pressure gradient in the liquid film is always much smaller than the
  pressure gradient in the gas film.\label{figcomp}}
\end{figure}

Typical interfaces profiles of the numerical solution of {\eqref{lub1}},
{\eqref{lub2}} and the vertical force balance {\eqref{eqmeanheight}} are seen in figure \ref{figcomp} for two different configurations~: on the right, the dynamics in the liquid film case ($\tmop{St}_1 \simeq 0.6$) appears truly different from the solid case shown on left ($\tmop{St}_1 = \infty$). We expect therefore the time of contact to be also different and to depend
on $\tmop{St}_1$. The present paper aims therefore at understanding how this contact time depends on the liquid film properties, focusing on the influence of $\tmop{St}_1$ on the drainage of the air film.

\subsection*{Drainage dynamics}
\begin{figure}[h]
  \includegraphics[width=8.7cm]{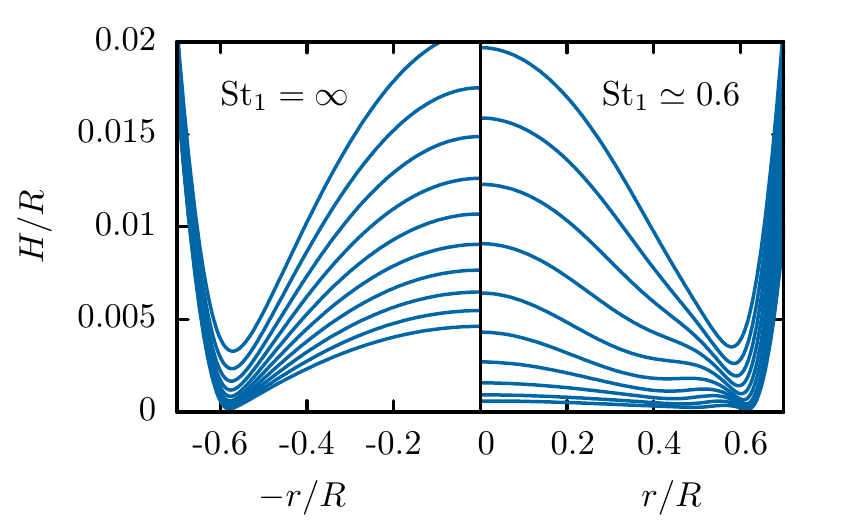}
  \caption{Successive plots of the dimensionless gas film height $H/R$ in the case of a solid film ($\tmop{St}_1 \rightarrow \infty$, left) and a liquid film ($\tmop{St}_1 = 0.6$, right), for $\tau=t\sqrt{g/R}=0.7\times2^{n}$ with $n \in [0,9]$. Parameters are : $\tmop{Bo} = 1.01$, $\tmop{St}_2 = 1.25 \times 10^{- 4}$. \label{figprofiles}}
\end{figure}

In that purpose, we investigate and clarify first the dimple drainage dynamics for these two characteristic cases $\tmop{St}_1 = \infty$ and $\tmop{St}_1 = 0.6 $. 
Figure \ref{figprofiles} presents the evolution in time of
the air gap $H/R$ as a function of $r/R$. As for figure \ref{figcomp}, the left part corresponds to the solid
substrate ($\tmop{St}_1 \rightarrow \infty$) and the right part to the
liquid film ($\tmop{St}_1 \simeq 0.6$). 
In both cases, the profiles correspond to $\tau=t\sqrt{g/R}=0.7\times2^{n}$ with $n \in [0,9]$ in dimensionless unit (see Materials and Methods).

After the dimple is formed, at short time, the drainage appears much faster in the
case of a liquid film underneath. This fact is closely related to the
continuity of tangential velocity across interface $1 / 2$ : in the solid
case, this velocity is zero, whereas it is finite in the case of a liquid
film. Therefore, viscous dissipation in the gas film at the tip of the dimple
(where $H$ is minimal) slows down more efficiently the dimple drainage, by contrast with the
liquid film case, where the gas can more easily escape from the dimple. Moreover, the liquid film deforms under the pressure
field for finite $\tmop{St}_1$ so that the drop shape underneath and the film thickness almost coincide, as it can be seen on figure \ref{figcomp} right.

\begin{figure}[h]
  \includegraphics[width=8.7cm]{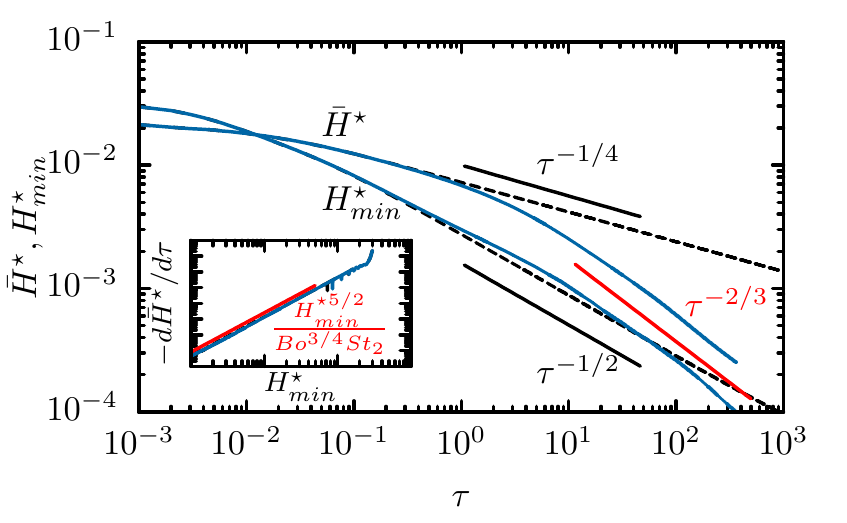}
  \caption{Dimensionless mean height of the dimple $\bar{H}^\star=\bar{H}/R$ and minimum height $H_{\min}^\star=H_{min}/R$ as a function of dimensionless time $\tau$ in the case of a liquid film ($\tmop{St}_1 = 0.6$, solid curves) and a solid film ($\tmop{St}_1 \rightarrow \infty$, dashed curves).
  Insert~: $-d\bar{H}^\star/d\tau$ as a function of $H_{min}^\star$ and the predicted law $H_{min}^{\star5/2} / \tmop{Bo}^{3/4} \tmop{St}_2$. Ranges are $[10^{-4},10^{-1}]$ in $x$ and $[10^{-7},10]$ in $y$. Solid case : black dashed curve; liquid film case : blue solid curve. 
  \label{figscalings}}
\end{figure}

Beside these qualitative observations, can we understand more quantitatively the dynamics of the dimple drainage in both cases?
Figure \ref{figscalings} shows the evolution with time for the two cases shown on figure \ref{figprofiles} ($\tmop{St}_1 = 0.6$ and $\tmop{St}_1 = \infty$) of the two relevant quantities to describe the dimple geometry, $\bar{H}(t)$ the dimple mean height, defined below, and the minimal air gap, at the neck, $H_{\min} (t)=H(r_{\min}(t),t)$, where $r_{\min}(t)$ is its radial neck position (which is only slightly varying with time). Different scalings with time are observed in the asymptotic limit $t \rightarrow \infty$: for the solid substrate case $\tmop{St}_1 = \infty$ (black dashed lines), we have $\bar{H} \sim t^{-1/4}$ and $H_{\min} \sim t^{-1/2}$, while for the liquid film $\tmop{St}_1 = 0.6$  (blue solid lines), the dynamics follows $\bar{H} \sim H_{\min} \sim t^{-2/3}$.

To explain these scalings, let us first start with the solid case limit, where {\eqref{lub2}} reduces to the well-known
lubrication equation :
\begin{equation}
  \frac{\partial H}{\partial t} = \frac{1}{3 \eta_2} 
  \frac{\partial}{\partial r} \left( H^3 \frac{\partial p_2}{\partial r}
  \right) \label{lubsol}
\end{equation}
for which the drainage dynamics has been already characterized~\cite{frankel1962dimpling,hartland1977model,lin1982thinning,hartland1994dimple}. We will recall the main
results here for the consistency of our paper and also because it gives the framework for solving the finite $\tmop{St}_1$ cases.
The interface geometry consists of a dimple of constant width, deflating
slowly, connected to a small gap region where viscous dissipation dominates,
located in $r_{\min}$, that we can take constant when we approach the drainage time. We are looking for radial and vertical scalings of this
small gap region, namely $\ell (t)$ and $H_{\min} (t)$ respectively. 
In the dimple, the gas flux $Q = - \frac{H^3}{3 \eta_2} \frac{\partial
p_2}{\partial r}$ can be neglected, leading to a uniform pressure $P_2$. This
is confirmed by figure \ref{figcomp}, where the red curve on the left (solid
case) corresponds to the pressure in the gas film. Integrating 
{\eqref{lap2}} twice according to $r$, where the gravity term can be neglected for small dimple, we obtain $h_2 (r)$, and therefore $H$,
assuming that $H (r_{\min}) \simeq 0$~:
\begin{equation}
	H (r, t) = \frac32 \frac{r_{min}^2-r^2}{r_{min}^2}\bar{H}(t) ,
\label{Hrt}
\end{equation}
where 
$$
\bar{H}(t) \equiv \frac{1}{r_{min}} \int_0^{r_{min}}{H} dr.
$$

The curvature of the small gap region
needs to be matched on its right to the curvature of the sessile top surface, yielding~:
\begin{equation}
  \frac{H_{\min}}{\ell^2} \sim  \frac{1}{\ell_c},
\label{scaling2}
\end{equation}
where $\ell_c=\sqrt{\gamma/\rho_3 g}$ is the capillary length. 
Moreover, the decrease per unit time of the dimple surface $S_d$ is equal to the leaking gas flux in the small
gap. Indeed, integrating {\eqref{lubsol}} between $r = 0$ and $r =
r_{\min}$, we get for the surface decrease :
	\begin{equation}
- \frac{d S_d}{d t}
		=- r_{min} \frac{d \bar{H}}{dt} 
		= - \frac{H_{\min}^3}{3
   \eta_2} \frac{\partial p_2}{\partial r}
   , 
		\label{scalingHmoy}
	\end{equation}
Remind that we do our analysis in 2D, but notice that the scalings would be unchanged in the 3D analysis since the self-similar dynamics is around a fixed radius $r_{min}$.
   In the thin gap region, the pressure gradient is large and given by equation \eqref{lap2}~:
   \begin{equation}
	   \frac{\partial p_2}{\partial r} \sim - \gamma \frac{\partial^3 H}{\partial r^3}
	   \sim - \gamma \frac{H_{min}}{\ell^3}.
	   \label{gradp2}
   \end{equation}
Moreover, balancing the pressure in the dimple with the drop weight and integrating the pressure gradient accross the neck gives the following relations~:
\begin{equation}
P_2 \sim   \frac{\rho_3 g R^2}{r_{min}} \;\;\;\; {\rm and} \;\; \;\; r_{min} \sim \tmop{Bo} \ell_c,
 \label{scalingP2}
\end{equation}
where we have used the fact that the scaling for the surface of our 2D drop is $S \sim R^2$, which is true for small to moderate Bond numbers.
Using \eqref{scaling2}, \eqref{scalingHmoy}, \eqref{gradp2} and \eqref{scalingP2}, we obtain~:
\begin{equation}
\frac{d \bar{H}}{dt} \sim - \frac{\gamma H_{\min}^4}{\eta_2 \ell^3 r_{min}} \label{scaling3}
\sim
 -\frac{\sqrt{g R}}{\tmop{Bo}^{3/4} \tmop{St}_2}\left( \frac{H_{min}}{R} \right)^{5/2} 
\end{equation}

So far, we have only used lubrication theory, and the matching of the curvature in $r_{min}$ to the outer curvature given by the top sessile drop. 
To close the system of equations, we need a relation between the neck and the inner region. For the solid film case, it can be done by matching the slopes between the dimple and the neck.
Using the dimple geometry \eqref{Hrt} and the length scales relevant in the neck, this matching yields in scaling~:
\begin{equation}
	\frac{H_{\min}}{\ell} \sim \frac{\bar{H}}{r_{min}}  \label{scaling1},
\end{equation}
a result that could be obtained using asymptotic matching between the two
regions {\cite{duchemin2005static}}. 
Finally, using \eqref{scaling2}, \eqref{scaling3} and \eqref{scaling1}, we obtain~:
\begin{equation}
	\ell \sim \frac{\bar{H}}{\tmop{Bo}},  
\quad 
H_{\min} \sim \frac{R}{\tmop{Bo}^{3/2}} \left( \frac{\bar{H}}{R} \right)^2,
\label{scalingsolidlength1}
\end{equation}
and
\begin{equation}
	\frac{d \bar{H}}{dt} 
	\sim
	-\frac{\sqrt{g R}}{\tmop{Bo}^{9/2} \tmop{St}_2}
	\left( \frac{\bar{H}}{R} \right)^5.
\label{scalingsolidlength2}
\end{equation}
Integrating \eqref{scalingsolidlength2} in time, we finally obtain the following long-time behaviours, already obtained for instance by {\cite{yiantsios1990}}:
\begin{eqnarray}
	\frac{\bar{H}}{R} \sim \tmop{Bo}^{9/8} \tmop{St}_2^{1/4} \tau^{- 1 / 4}, \\
  \frac{H_{\min}}{R} \sim  \tmop{Bo}^{3/4} \tmop{St}_2^{1/2} \tau^{-1/2}, \\ 
 \frac{\ell}{R} \sim \tmop{Bo}^{1/8} \tmop{St}_2^{1/4} \tau^{- 1 / 4}
 \label{scalingsolid},
\end{eqnarray}
where $\tau=t \sqrt{g/R}$ is the dimensionless time.
These scalings in $\tau$ are confirmed by the numerical solution of 
{\eqref{lub1}} and {\eqref{lub2}}, as seen in figure \ref{figscalings}, where
the dashed curves correspond to $\bar{H}/R$ and $H_{\min}/R$ in the limit $\tmop{St}_1 \rightarrow \infty$. 

In the liquid case, we have to question the dominant balance in 
{\eqref{lub2}}. This balance is actually the same as for the solid case, for
two reasons. The first one is that $| \partial p_2 / \partial r | \gg |
\partial p_1 / \partial r |$ in the small gap region, as confirmed in figure
\ref{figcomp}. This is mainly due to surface tension that damps the
high-curvature regions on the interface 1/2, and therefore smooths the
pressure gradient. The other reason is that in our configuration, and in the
experiments described in {\cite{lo2017}}, $\tmop{St}_2 \ll \tmop{St}_1$, such
that the first term in the right-hand-side of \eqref{lub2}
is bigger than the third term, namely :
\[ \left| \frac{H^3}{3 \eta_2} \frac{\partial p_2}{\partial r} \right|
   \gg \left| \frac{h_1 H^2}{\eta_1} \frac{\partial p_2}{\partial r}
   \right| , \]
 during our numerics. This condition gives in fact the validity range 
 $$  H \gg \frac{\eta_2}{\eta_1} h_1, \quad {\rm or} \quad \frac{H}{h_1} \gg \frac{\tmop{St}_2}{\tmop{St}_1} $$
 for the scaling deduced below, suggesting that another scaling should be observed for smaller neck thickness than those
 computed numerically.
Finally the dominant balance in the lubrication {\eqref{lub2}} is
still consistent with {\eqref{lubsol}}. So, why are the self-similar
scaling observed in figure \ref{figscalings} so different between the two
cases? The answer lies in the fact that the scalings of $H_{\min}$ and $\bar{H}$
are no longer related through {\eqref{scaling1}}, which is due to the matching between the parabolic dimple and the neck.
For the liquid film case, the dimple geometry is very different, showing a flater structure (figure \ref{figprofiles} right), and our numerics suggests that the dynamics of
$H_{\min}$ and $\bar{H}$ are similar~:
\begin{equation}
	H_{\min} \sim \beta \bar{H} \label{scaling1b}
\end{equation}
up to a prefactor $\beta$ that is a priori very small and depends on all the control parameters~: Bo, St$_1$, St$_2$, and $\alpha$. Using \eqref{scaling2} and \eqref{scaling3}, which are still valid in the liquid film case (since the curvature of the drop interface is much larger than the curvature of the film interface), and {\eqref{scaling1b}}, we obtain :
\begin{eqnarray}
	\frac{\bar{H}}{R} \sim 
	\tmop{Bo}^{1/2} \tmop{St}_2^{2/3} \beta^{-5/3} \tau^{- 2 / 3}, \\
	\frac{H_{min}}{R} \sim 
	\tmop{Bo}^{1/2} \tmop{St}_2^{2/3} \beta^{-2/3} \tau^{- 2 / 3}, \\
	\frac{\ell}{R} \sim 
	\tmop{St}_2^{1/3} \beta^{-1/3} \tau^{- 1 / 3}.
	\label{scalingliquid}
\end{eqnarray}
These scalings in $\tau$ are again confirmed by numerical simulations, as seen in figure
\ref{figscalings}. The agreement, although very good, is not perfect because
the dominant balance used in {\eqref{lub2}} is only an approximation here.

\begin{figure}[h]
  \includegraphics[width=8.7cm]{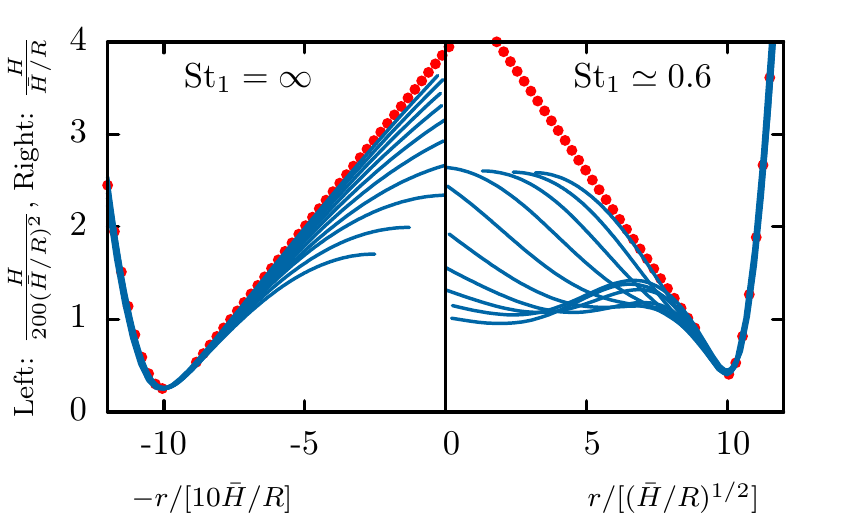}
  \caption{Same profiles as in figure \ref{figprofiles} rescaled according to $[10 \bar{H}/R,200 (\bar{H}/R)^2]$ (Solid case, left), $[(\bar{H}/R)^{1/2},\bar{H}/R]$ (Liquid case, right), and shifted horizontally in order for the neck to be at the same position for all profiles. The linear and quadratic behaviors corresponding to the critical solution of \eqref{current} are represented as red dots.
\label{figprofiles_rescaled}}
\end{figure}

Moreover, our approaches to determine the scalings assume somehow a self similar shape of the interface near the neck in the form:
\begin{equation}
H(r,t) \sim H_{\min} (t) F \left(\frac{r-r_{\min}}{\ell(t)} \right).
\label{eqself}
\end{equation}
Such self-similar character of the successive film thicknesses $H$, \eqref{eqself}, can be verified. In the solid case, from \eqref{scaling2} and \eqref{scaling1}, we obtain $\ell \sim \bar{H}$ and $H_{min} \sim \bar{H}^2$, while in the liquid film case, from equations \eqref{scaling2} and \eqref{scaling1b}, we get $\ell \sim \bar{H}^{1/2}$ and $H_{min} \sim \bar{H}$. Therefore, rescaling the radial and vertical coordinates of the interface around the neck according to these scalings allows to check for spatial self-similarity, as seen in figure \ref{figprofiles_rescaled}. The interface profiles rescale nicely near the neck for each case, while it does not at all in the dimple region.
Remarkably, plugging this sef-similar ansatz \eqref{eqself} into the lubrication equation (\ref{lubsol}) does not lead to the balance between the left and the right hand side of the equation for the scalings obtained above, showing in fact that the evolution of $H_{min}$ is subdominant versus the mass flux terms. Indeed, neglecting the time-dependent and gravity terms in front of surface tension, we obtain the so-called ``current equation''~:
\begin{equation}
	F^3 F_{\xi \xi \xi} = C,
	\label{current}
\end{equation}
where $\xi=(r-r_{min})/\ell(t)$ and $C$ is a constant. This equation is studied in great detail in \cite{lamstaes}, where we learn that the selected solution to this equation is usually the critical case, for which growth is quadratic on one side, and linear on the other. This remark should be valid in the neck region, in both the liquid and solid film cases. 
We have verified this fact on the rescaled profiles shown in figure \ref{figprofiles_rescaled}, and it turns out that in both cases, the self-similar shape behaves like $\xi$ towards the inner region (dimple) and $\xi^2$ towards the outer region (sessile drop), as indicated on the figure with red dots. This result is also consistant with the right and left matchings used to obtain the self-similarity exponents. 

\subsection*{Contact time}

\begin{figure}[h]
  \includegraphics[width=8.7cm]{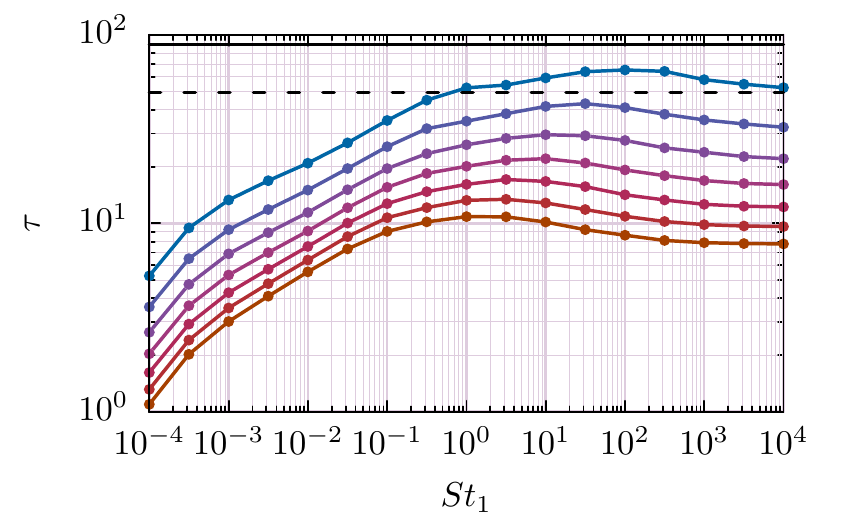}
  \caption{Dimensionless time $\tau$ for $H_{\min}/R$ to reach a given threshold (from $4 \times
  10^{- 4}$ for blue to $10^{- 3}$ for brown, with an increment of $10^{-4}$). The black dashed and black
  solid lines correspond to $4 \times 10^{- 4}$ and $3 \times 10^{- 4}$
  respectively, in the solid case. \label{figtau}}
\end{figure}

An interesting output of the present study, relevant to the recent
experimental observations {\cite{lo2017}}, is the time for the gas film to
reach a thickness small enough such that coalescence can occur. Figure
\ref{figtau} shows the dimensionless time it takes for the minimum gas film
dimensionless height $H_{\min}/R$ to reach a given value, as a function of the Stokes number
$\tmop{St}_1$. The curves correspond to decreasing thresholds (from
$10^{- 3}$ to $4 \times 10^{- 4}$ from bottom to top). Finally, the two
lines correspond to the solid case for a threshold $4 \times 10^{- 4}$
(dashed) and $3 \times 10^{- 4}$ (solid). Although the fluid cases exhibit a
local maximum, for $1 \lesssim \tmop{St}_1 \lesssim 100$, even higher than
the solid curve ($\tmop{St}_1 \rightarrow \infty$) for the minimal threshold
($4 \times 10^{- 4}$), we suspect that this local maximum disappears as the threshold decreases. Indeed,
the lower threshold limit $3 \times 10^{- 4}$ in the solid case is well above
the last curve for the liquid case. However, keeping in mind that we want to
compare this length to a rupture critical gas layer thickness it is interesting to
notice that there could be an optimum value of the Stokes number $\tmop{St}_1$
for which the coalescence time is maximum. Although we do not have a definitive explanation for the existence of this maximum time to reach a given threshold, we would like to argue that this behavior comes from the competition between the gas film drainage and the squeezing of the liquid film. Indeed, in this intermediate regime the liquid film still deforms highly so that the gas film drainage is affected by the liquid film geometry.

In order to estimate the drainage time before the film rupture, we need to estimate the typical length at which the physical mechanism for the rupture is pertinent. For the deposition on a liquid film, Van der Waals forces~\cite{zhang1999similarity} or non-continuum effects~\cite{DJ12} become relevant typically around and below $100$ nm, leading for a typical drop size of $1$ mm to a dimensionless threshold of $10^{-4}$ of the order of the smallest threshold investigated numerically on figure \ref{figtau}. 
On the other hand, for solid deposition, the roughness of the substrate is often the dominant mechanism for the film rupture. Taking a smooth substrate of roughness $1 \mu$m, we obtain a dimensionless threshold of $10^{-3}$. These estimates might explain while, although the drainage on a thin liquid film is faster than on a solid substrate, the coalescence time can be smaller for solid substrate, as observed eventually in experiments~\cite{lo2017}.

\subsection*{Conclusion}

In this study, we have been investigating the drainage dynamics of a thin viscous liquid film squeezed by a millimetric droplet under gravity. We have shown a significant drainage speed-up when a liquid film covers the solid substrate, even if it is very viscous (Figures \ref{figscalings} and \ref{figtau}). The drainage time can therefore vary of few orders of magnitude as the viscosity of the liquid film changes. 
Nevertheless, the final coalescence time remains dependent of the final physical mechanism of the film rupture and our approach allows for a quantitative prediction of the coalescence time when the threshold lenghtscale is determined.
Although our self-similar predictions \eqref{eqself} are probably difficult to measure experimentally, it would be very interesting to measure rigorously the coalescence time when varying the viscosity and the thickness of the liquid film. Moreover, we leave as a perspective of the present work parametric studies consisting in varying systematically these parameters, and the volume of the drop, through the Bond number.

\matmethods{

In order to solve numerically {\eqref{lub1}} and {\eqref{lub2}}, we
need boundary conditions in $r = 0$ and $r = R$. In $r = 0$, we use symmetric
boundary conditions ($h_1' = h_1''' = 0$, $H' = H''' = 0$). In $r = R$, $h_1$
is matched onto an outer static meniscus, $H'$ is matched to its value given
by the sessile top surface. The last condition for $H$ is given by the
vertical force balance :
\begin{equation}
  \rho_3 \mathcal{S} g = \int_0^R p_2 \tmop{dr} = p_0 (t) R
  - \rho_3 g R \overline{h_2} - \gamma \frac{\partial h_2}{\partial r} (R),
\end{equation}
where we have used equation {\eqref{lap2}}, $\mathcal{S}$ is the surface of
half of the drop (a surface in our 2D numerics in fact), and
\[ \overline{h_2} = \frac{1}{R}\int_0^R h_2 \tmop{dr}. \]
$p_2$ is supposed to relax towards $0$ in $r = R$ :
\[ 0 = p_0  - \rho_3 g h_2 (R) - \gamma \frac{\partial^2 h_2}{\partial
   r^2} (R), \]
giving $p_0$. Finally, we get an integral equation for $h_2$ :
\begin{equation}
	\frac{\mathcal{S}}{R} = h_2 (R) + \frac{\gamma}{\rho_3 g} \frac{\partial^2
	h_2}{\partial r^2} (R) - \frac{1}{R}\int_0^R h_2 \tmop{dr} - \frac{\gamma}{\rho_3 g R}
  \frac{\partial h_2}{\partial r} (R) \label{eqmeanheight}
\end{equation}
which is solved at each timestep together with equations {\eqref{lub1}} and
{\eqref{lub2}}.

Equations {\eqref{lub1}}, {\eqref{lub2}}, \eqref{lap2}, \eqref{lap1} and
{\eqref{eqmeanheight}} are made dimensionless according to the length scale $R$, velocity scale $\sqrt{g R}$ and pressure scale $\rho_3 V^2 = \rho_3 g R$. $R$ is chosen to be the radial distance at which the slope of the sessile top surface is equal to $1$ (figure \ref{sketch}). 
Plugging {\eqref{lap2}} and {\eqref{lap1}} into 
{\eqref{lub1}} and {\eqref{lub2}}, we obtain two non-dimensional equations for
$h_1^\star=h_1/R$ and $H^\star=H/R$ involving spatial derivatives of $h_1^\star$ and $H^\star$ up to the fourth derivative.
We use a second-order semi-implicit finite-difference
method, treating implicitly the spatial derivatives in the right-hand-side of
{\eqref{lub1}}, {\eqref{lub2}}. Doing so, we get rid of the stiffness
of these equations, coming from the high-order spatial derivatives~\cite{salez2012numerical,Duchemin2014}.

\eqref{lub1} and \eqref{lub2} read respectively, after dropping the $^\star$~: 

\begin{eqnarray*}
  \ds\frac{\partial h_1}{\partial \tau}  & = & h_2' \left( \ds\frac{- h_1^2
  h_1'}{\tmop{St}_1} - \frac{6 h_1 h_1' H + 3 h_1^2 H'}{6 \tmop{St}_1} \right)\\
  &  & + h_2'' \left( \ds\frac{- h_1^3}{3 \tmop{St}_1} - \frac{h_1^2 H}{2
  \tmop{St}_1} \right)\\
  &  & + h_2''' \left( \ds\frac{- 2 h_1^2 h_1'}{\tmop{St}_1 \tmop{Bo}} - \frac{6
  h_1 h_1' H + 3 h_1^2 H'}{6 \tmop{St}_1 \tmop{Bo}} \right)\\
  &  & + h_2'''' \left( \ds\frac{- 2 h_1^3}{3 \tmop{St}_1 \tmop{Bo}} -
  \frac{h_1^2 H}{2 \tmop{St}_1 \tmop{Bo}} \right)\\
  &  & + H''' \left( \ds\frac{h_1^2 h_1'}{\tmop{St}_1 \tmop{Bo}} \right)\\
  &  & + H'''' \left( \ds\frac{h_1^3}{3 \tmop{St}_1 \tmop{Bo}} \right), 
\end{eqnarray*}

\begin{eqnarray*}
     {\ds\frac{\partial H}{\partial \tau}} & = & h_2' \left( \ds\frac{-
     H^2 H'}{\tmop{St}_2} - \frac{2 h_1 h_1' H + h_1^2 H'}{2 \tmop{St}_1} -
     {\frac{2 h_1' H^2 + 4 h_1 H H'}{2 \tmop{St}_1}} \right)\\
     &  & + h_2''  \left( \ds\frac{- H^3}{3 \tmop{St}_2} - \frac{h_1^2 H}{2
     \tmop{St}_1} - {\frac{h_1 H^2}{\tmop{St}_1}} \right)\\
     &  & + h_2''' \left( \ds\frac{- H^2 H'}{\tmop{St}_2 \tmop{Bo}} -
     {\frac{2 h_1 h_1' H + h_1^2 H'}{\tmop{St}_1 \tmop{Bo}}} -
     {\frac{2 h_1' H^2 + 4 h_1 H H'}{2 \tmop{St}_1 \tmop{Bo}}}
     \right)\\
     &  & + h_2'''' \left( \ds\frac{- H^3}{3 \tmop{St}_2 \tmop{Bo}} -
     {\frac{h_1^2 H}{\tmop{St}_1 \tmop{Bo}}} -
     {\frac{h_1 H^2}{\tmop{St}_1 \tmop{Bo}}} \right)\\
     &  & + H''' \left( {\ds\frac{2 h_1 h_1' H + h_1^2 H'}{2
     \tmop{St}_1 \tmop{Bo}}} \right)\\
     &  & + H'''' \left( {\ds\frac{h_1^2 H}{2 \tmop{St}_1
     \tmop{Bo}}} \right),
\end{eqnarray*}
where $h_2=h_1+H$ and $'$ denotes differentiation according to $r$.
For all the computations presented in this article, the dimensionless spatial domain $r \in [0,1]$ is discretized into 201 intervals and the timestep is $\delta t = 10^{-3}$.

All the data included in the present article and the code used to solve the equations and obtain these data are available at \href{http://dx.doi.org/10.4121/uuid:c09719c0-8404-43e1-b1f1-1833ded0e87a}{http://dx.doi.org/10.4121/uuid:c09719c0-8404-43e1-b1f1-1833ded0e87a}. 

}

\showmatmethods{} 

\acknow{Please include your acknowledgments here, set in a single paragraph. Please do not include any acknowledgments in the Supporting Information, or anywhere else in the manuscript.}


\bibliography{../coalescence}

\end{document}